\documentclass[12pt]{article}
\pdfoutput=1
\usepackage{hyperref}
\usepackage{graphicx}
\usepackage{graphics}
\usepackage{dsfont}
\usepackage{epsfig}
\usepackage{amsmath,amssymb,amsthm,amscd}
\usepackage{simpler-wick}

\newcommand{\be}{\begin{equation}}
\newcommand{\ee}{\end{equation}}
\newcommand{\ba}{\begin{aligned}}
\newcommand{\ea}{\end{aligned}}

\numberwithin{equation}{section}

\begin{document}
\begin{titlepage}

\rightline{USTC-ICTS/PCFT-23-20}

\vskip 3 cm

\centerline{\Large 
\bf  
Thermal Entropy in   } 
\vskip 0.2 cm
\centerline{\Large 
\bf  
 Calabi-Yau Quantum Mechanics   }

\vskip 0.5 cm

\renewcommand{\thefootnote}{\fnsymbol{footnote}}
\vskip 30pt \centerline{ {\large \rm 
Min-xin Huang\footnote{minxin@ustc.edu.cn}  
} } \vskip .5cm  \vskip 20pt 

\begin{center}
{Interdisciplinary Center for Theoretical Study,  \\ \vskip 0.1cm  University of Science and Technology of China,  Hefei, Anhui 230026, China} 
 \\ \vskip 0.3 cm
{Peng Huanwu Center for Fundamental Theory,  \\ \vskip 0.1cm  Hefei, Anhui 230026, China} 
\end{center}

\setcounter{footnote}{0}
\renewcommand{\thefootnote}{\arabic{footnote}}
\vskip 40pt
\begin{abstract}

We consider the von Neumann entropy of a thermal mixed state  in quantum systems derived from mirror curves, where the kinetic terms are exponential functions of the momentum operators. Using the mathematical results on the asymptotics of the energy eigenvalues, we compute the asymptotic entropy in high temperature limit and compare with that of the conventional models.  We discuss the connections with some folklores in quantum gravity, particularly on the finiteness of entropy.

\end{abstract}

\end{titlepage}
\vfill \eject


\newpage

\baselineskip=16pt

\tableofcontents

\section{Introduction}

Entropy has played a significant role in recent developments in our understandings of holography and quantum gravity. Since the seminal proposal of holographic entanglement entropy in \cite{Ryu:2006bv}, there have been many generalizations and applications. In particular, a generalized version of the entropy may follow the Page curve of the black hole entropy during the Hawking radiation process, as reviewed in  \cite{Almheiri:2020cfm}. These significant progress appear to be getting close toward a consensus on the resolution of the famous black hole information paradox. 

The thermal mixed state in a quantum system can be obtained as the reduced state by tracing over one party in a thermofield double (TFD) state, which is a pure quantum state entangled between two similar systems. The von Neumann entropy of the thermal state is the entanglement entropy of the TFD state. The use of TFD states has been instrumental in our understandings of holography with two boundaries e.g. in \cite{Maldacena:2001kr}, as well as in the study of von Neumann algebra e.g.  in \cite{Witten:2021jzq}. Thus it is well motivated to consider the entropy of a thermal mixed state in various interesting quantum systems.

We will consider quantum systems derived from mirror curves, where the kinetic terms are exponential functions of the momentum operators, with both exponential and polynomial potentials for the position operators respectively in Sections \ref{CYsection} and \ref{4dsection}. The Hamiltonians derived from toric Calabi-Yau geometries are exponential function of the position and momentum operators. The most well studied models are  the local $\mathbb{P}^1\times\mathbb{P}^1$ model and the local $\mathbb{P}^2$ model
	\be\ba  \label{CYHamiltonian}
		\hat{H}&=e^{\hat{x}}+e^{-\hat{x}}+e^{\hat{p}}+e^{-\hat{p}}, ~~~~ \mathbb{P}^1\times\mathbb{P}^1 ~\textrm{model},\\
		\hat{H}&=e^{\hat{x}}+e^{\hat{p}}+e^{-\hat{x}-\hat{p}}, ~~~~~~~~~  \mathbb{P}^2 ~\textrm{model} .\\
	\ea\ee 
The perturbative quantization conditions are given by the Nekrasov-Shatashvili limit of refined topological string theory \cite{Nekrasov:2009rc, Aganagic:2011mi}. Furthermore, based on some earlier works on numerical calculations of the quantum spectrum \cite{Kallen:2013qla, Huang:2014eha},  the exact quantization conditions including all non-perturbative effects were conjectured in \cite{Grassi:2014zfa, Wang:2015wdy}, where the two seemingly different proposals turned out to be related  by the blowup equations \cite{Grassi:2016nnt}. Now known as the TS/ST (Topological String/Spectral Theory) correspondence, it has attracted the attentions of many mathematicians as well. For example,  some promising partial results toward a proof of the conjecture are obtained in the recent papers \cite{Doran:2021hcy, Gavrylenko:2023ewx}. In particular, some nice mathematical results on the asymptotics of the energy eigenvalues were proven in \cite{LST2016, LST2019}, which we will use to study the asymptotic entropy in high temperature limit. 

As a warm up exercise and a standard to compare, we first consider the simple case of a harmonic oscillator. It is well known that one can couple two harmonic oscillators together with a quadratic interaction, then the  ground state of the combined system is a TFD state \cite{Srednicki:1993im}. Some strategies to build TFD states in more general systems were studied e.g. in \cite{Cottrell:2018ash}. For a harmonic oscillator 
\be \label{harmonic}
\hat{H} =\frac{\hat{p}^2}{2} +\frac{ \omega^2  \hat{x}^2}{2} ,
\ee
it is not difficult compute the von Neumann entropy of a thermal state 
\be 
S = \frac{\beta \omega}{e^{\beta \omega}-1} -\log(1-e^{-\beta \omega}),
\ee
where we have set $\hbar=1$ and the parameter $\beta \equiv \frac{1}{k_B T}$ is defined in terms of the Boltzmann constant and the temperature. At zero temperature, the thermal state is just the pure ground state so the entropy is zero. We will instead be interested in the high temperature limit where the probability is more evenly distributed between all excited states, so the entropy should be maximized. In this simple case the asymptotic behavior is 
\be \label{asym1}
S\sim  \log(T) , ~~~~~ T\sim \infty .
\ee

More generally, let $0<\lambda_1\leq \lambda_2 \leq  \cdots$ denote the eigenvalues of a Hamiltonian. We will use the number of states defined by 
\be 
N(\lambda)  = \#\{ j\in \mathbb{N}: \lambda_j < \lambda \}.
\ee
For a thermal state, the probability at an excited state is $p_i = \frac{1}{Z} e^{-\beta \lambda_i}$, where $Z=\sum_{i} e^{-\beta \lambda_i}$ is the partition function. The von Neumann entropy can be written as 
\be \label{S1.5}
S= - \sum_{i=1}^{\infty} p_i \log(p_i) = \log(Z) + \frac{\beta}{Z} \sum_i \lambda_i e^{-\beta \lambda_i} =\log(Z) + T\partial_T \log(Z).
\ee
The partition function can be computed by 
\be   \label{laplace}
Z=\int_0^\infty e^{-\beta \lambda} d N(\lambda).  
\ee
To compute the asymptotic behavior of the entropy, we only need the asymptotic behavior of $N(\lambda)$. For the harmonic oscillator, the number of states goes like $N(\lambda)\sim \lambda$, so we have $Z\sim T$ in the high temperature limit and the first term in (\ref{S1.5}) dominates. We recover the result (\ref{asym1}).

For a non-relativistic quantum system with a standard kinetic term as in (\ref{harmonic}) but a general polynomial potential, it is expected that the number of states grows  like a power law $N(\lambda)\sim \lambda^c$ with a positive number $c$. In this case it is easy to see that the partition function scales like $Z\sim T^c$ and the  first term in (\ref{S1.5}) dominates. So the thermal entropy also scales like $S\sim \log(T)$ in the high temperature limit.

\section{Calabi-Yau Quantum Mechanics} \label{CYsection}
As we mentioned, the simplest examples of such quantum models are (\ref{CYHamiltonian}).  Some more general models were considered in \cite{LST2016}, including a deformation the $\mathbb{P}^1\times\mathbb{P}^1$ model, and a more general operator $e^{-m \hat{x}- n \hat{p}}$ for arbitrary natural numbers $m,n$ in the $\mathbb{P}^2$ model. A common feature is that the number of states grows like 
\be 
N(\lambda) \sim \log^2(\lambda), ~~~~ \lambda \sim \infty
\ee
where we have neglected  the coefficient factor which depends on specific models. We see that the logarithmic growth is much slower than the linear growth of the harmonic oscillator. As a result, the inverse Hamiltonian $\hat{H}^{-1}$ is a trace class operator for the Calabi-Yau models, which was first proven by other method in \cite{Kashaev:2015kha}, while it is not for the harmonic oscillator.

\newcommand{\MeijerG}[7]{G ^ {#1 , #2 } _{#3 , #4} \left( \begin{matrix} #5 \\ #6 \end{matrix} ~\middle\vert ~ #7 \right) }

In the high temperature limit, the partition function is 
\be  
Z \sim f(\beta)\equiv \int_1^\infty e^{-\beta \lambda} \frac{\log(\lambda)}{\lambda}  d\lambda  . 
\ee
The integral  $f(\beta)$ is convergent for  $\Re(\beta)>0$ and can be analytically continued to the complex plane with potential branch cuts. Using the \textit{Mathematica} program, we find that it can be written in terms of a Meijer G-function and  it has the following series expansion around $\beta\sim 0$
\be \label{MeijerG}
\ba
f(\beta) &= \MeijerG{3}{0}{2}{3}{1, 1}{0, 0, 0}{\beta}  \\
&= \frac{1}{2}\log^2(\beta)+ \gamma \log(\beta) + \frac{\gamma^2}{2}  + \frac{\pi^2}{12}  -\beta + \frac{\beta^2}{8} - \frac{\beta^3}{54} +\mathcal{O}(\beta^4),
\ea
\ee
where $\gamma$ is the Euler-Mascheroni constant. Here the Meijer G-function is a very general function including as particular cases many known special functions, such as generalized hypergeometric functions,  Bessel functions. 

Motivated by the series expansion (\ref{MeijerG}), we can provide a more elementary computation of the integral by the use of differential equation  
\be
\ba
 \frac{d}{d\beta} [ \beta f^{\prime}(\beta) ] & = \int_1^\infty e^{-\beta \lambda} (-1+\beta \lambda) \log(\lambda)  d\lambda  \\
 &= -\frac{1}{\beta}  \int_1^\infty d [e^{-\beta\lambda} (1+\beta \lambda \log(\lambda))] = \frac{e^{-\beta}}{\beta}. 
 \ea
\ee
Expanding the simple function $ \frac{e^{-\beta}}{\beta}$ and integrating twice, we can recover the series expansion (\ref{MeijerG}) except for the integration constant terms $ \gamma \log(\beta) + \frac{\gamma^2}{2}  + \frac{\pi^2}{12}$.

We can further give an alternative simple elementary derivation of the leading scaling behavior. Suppose $0<\beta<1$, the integral can be dissected into two parts 
\be 
f(\beta)   = \int_1^{\frac{1}{\beta}} e^{-\beta \lambda} \frac{\log(\lambda)}{\lambda}  d\lambda + \int_{\frac{1}{\beta}}^{\infty}  e^{-\beta \lambda} \frac{\log(\lambda)}{\lambda}  d\lambda . 
\ee
The second part will be subdominant and  is evaluated as 
\be 
 \int_{\frac{1}{\beta}}^{\infty}  e^{-\beta \lambda} \frac{\log(\lambda)}{\lambda}  d\lambda = 
  \int_{1}^{\infty}  e^{-\lambda} \frac{\log(\lambda) - \log(\beta)}{\lambda}  d\lambda \sim  \log(T) .
\ee
The first part can be estimated as
\be \ba
& \int_1^{\frac{1}{\beta}} e^{-\beta \lambda} \frac{\log(\lambda)}{\lambda}  d\lambda  > e^{-1} \int_1^{\frac{1}{\beta}} \frac{\log(\lambda)}{\lambda}  d\lambda = \frac{1}{2e} \log^2(\beta),  \\
& \int_1^{\frac{1}{\beta}} e^{-\beta \lambda} \frac{\log(\lambda)}{\lambda}  d\lambda < \int_1^{\frac{1}{\beta}} \frac{\log(\lambda)}{\lambda}  d\lambda =   \frac{1}{2}  \log^2(\beta). 
\ea \ee
Without concerning about the constant factor, we confirm the leading scaling behavior $f(\beta)\sim \log^2(\beta)$ in this way. The precise factor is actually $\frac{1}{2}$ from the series expansion for the Meijer G-function (\ref{MeijerG}), coinciding with the upper bound in the above estimates.

So we find the leading asymptotic behavior of the partition function 
\be  \label{scalingZ}
Z\sim \log^2(T), ~~~~ T\sim \infty. 
\ee
As in the harmonic oscillator case, the first term in (\ref{S1.5}) dominates and we have 
\be  \label{CYscaling}
S\sim \log(Z) \sim \log(\log(T)), ~~~~ T\sim \infty. 
\ee
We see that the entropy grows much slower than the harmonic oscillator in the high temperature limit.

\section{Difference Operators with Polynomial Potentials} \label{4dsection}
In this section we consider the case of polynomial potentials, e.g. 
\be
\hat{H} =  e^{\hat{p}}+e^{-\hat{p}} +W(\hat{x}),
\ee
where $W(x) = x^{2N} + \cdots $ is chosen to be an even degree $2N$ polynomial, so that we have a confining potential and an infinite discrete spectrum. This class of Hamiltonians comes from the quantizations the 4d Seiberg-Witten curves, which are obtained in a special scaling limit of the geometrically engineering Calabi-Yau mirror curves. The exact quantization condition in this case can be obtained from the 4d Nekrasov partition function of Seiberg-Witten theories and was studied in \cite{Grassi:2018bci}. The mathematical result \cite{LST2019} for the asymptotics of the number of states is 
\be 
N(\lambda)  \sim \lambda^{\frac{1}{2N}} \log(\lambda) , ~~~~ \lambda \sim \infty . 
\ee
The growth is faster than the Calabi-Yau models in the previous section, but slower than the standard harmonic oscillator. The inverse Hamiltonian $\hat{H}^{-1}$ is still a trace class operator. 

Using the similar method as the previous section, we find the asymptotic behavior of the partition function is $Z\sim T^{\frac{1}{2N}} \log(T)$ and check that the  first term in (\ref{S1.5}) dominates. So the asymptotics of the entropy is 
\be  \label{polygrowth}
S\sim \log(Z) \sim \log(T) , ~~~~~ T\sim \infty. 
\ee  
So the scaling behavior is actually the same as the standard harmonic oscillator.

\section{No Finite Bound for Entropy}  \label{finitesection}

As a related question to the discussions in the next section, we consider whether it is possible to have a finite upper bound for thermal entropy in the infinite temperature limit $T\rightarrow \infty$ for some quantum systems. Here we show that under reasonable assumptions this is impossible. We assume that the Hilbert space is infinite dimensional where the energy eigenvalues can be shifted to be all positive. Further assume the sum in the partition function is convergent for any finite temperature, so the partition function $Z(T)$ is well defined for any $T$, i.e. there is no exponential growth of the number of states $N(\lambda)$ as in string theory. It is easy to check by taking derivative that both $Z(T)$ and $S(T)$ are  monotonically increasing functions of $T$. Using (\ref{S1.5}), we find 
\be
S(T) = \log(Z) + T\partial_T \log(Z)  >\log(Z). 
\ee
Since  the Hilbert space is infinite dimensional, the partition function $Z(T)$ tends to infinity as $T\rightarrow \infty$, therefore the entropy also tends to infinity and there is no finite upper bound.

\section{Discussions} 

We have found that the thermal entropy of Calabi-Yau quantum mechanics (\ref{CYscaling}) grew much slower than that of the standard harmonic oscillator in the high temperature limit, while the case of polynomial potential had the same growth (\ref{polygrowth}). Although we focus on simple quantum systems, our study may provide some useful experience for relevant questions in quantum gravity. In particular, in the context of the influential Swampland Program \cite{Vafa:2005ui}, some recent works have studied the species scale, the emergence string proposal and their thermodynamics, see e.g. \cite{Dvali:2007hz, Lee:2019wij, Cribiori:2023ffn}. In certain limit e.g. near boundary of the moduli space, a tower of infinite number of string states may become light. Another way to explore such emergence is the high temperature limit considered here, where the highly excited states become equally probable. Of course, it is well known that string theory has a Hagedorn temperature, inverse proportional to the string length, where the partition function diverges. In those contexts, the high temperature limit should probably mean a  temperature approaching the Hagedorn temperature. Our study may provide some useful techniques to understand the asymptotic behavior near the emergence. 

A folklore of quantum gravity is the finiteness of entropy, in contrast to its divergence in generic calculations in quantum field theory. An important source of motivation comes from the finite horizon area of de Sitter space, which appears to be the current state of our universe. See e.g. \cite{Banks:2000fe, Balasubramanian:2001rb, Witten:2001kn} for earlier discussions related to de Sitter space and e.g. \cite{Das:2022nxo, Dabholkar:2023ows} for some recent discussions.  Of course, the dimension of Hilbert space is infinite in perturbative string theory but this is not necessarily in conflict with the folklore. In a countable (separable) infinite dimensional Hilbert space, the von Neumann entropy of a mixed state of trace class would be generically still finite except in some very contrived circumstances. For example, for a probability distribution that scales as power law $p_n\sim n^{-\alpha}$ among an orthogonal basis of states, the convergence of the sum $\sum_{n=1}^{\infty} p_n$ is equivalent to $\alpha>1$, in which case the von Neumann entropy $-\sum_{n=1}^{\infty} p_n\log(p_n)$ is also finite. We may consider a more contrived probability distribution $p_n \sim \frac{1}{n\log^{\alpha}(n)}$. For $\alpha\leq 1$ the sum $\sum_{n=1}^{\infty} p_n$ is divergent, while for $\alpha> 2$ both sums $\sum_{n=1}^{\infty} p_n$ and $-\sum_{n=1}^{\infty} p_n\log(p_n)$ are convergent. So in this case in a limited range $1< \alpha \leq 2$ we can have a probability distribution where the entropy is infinite. 

In the quantum models we studied, the entropy is indeed finite at a finite temperature, but tends to infinity in the infinite temperature limit.   A similar situation appeared in our earlier study of the entropy of Berenstein-Maldacena-Nastase (BMN) strings  \cite{Huang:2019uue}. In that case, the pp-wave spacetime background is infinitely curved, the strings become effectively infinitely long and tensionless with degenerate spectra, so the Hagedorn temperature is zero. Instead, a real non-negative genus counting parameter in the dual BMN double scaling limit becomes the effective string coupling $g$, playing a similar role of temperature as in the current context. Due to the structure of dual free CFT correlators, we are only accessing a countable infinite dimensional subspace of the whole Hilbert space of excited string states. It was found that at finite coupling $g$, the entropy is indeed also finite, while it is naively expected that as $g\rightarrow \infty$, the probability would be evenly distributed among the infinite dimensional Hilbert subspace, so the entropy should likely tend to infinity, which by itself does not seem to violate any fundamental principle of quantum gravity. Nevertheless, it would be a pleasant surprise if it turns out that the entropy of BMN strings in \cite{Huang:2019uue} does have a finite upper bound as $g\rightarrow \infty$, strongly confirming a folklore of quantum gravity in an implausible fashion. Such a bound may be related to the entropy of our current universe, thus could provide a natural estimate of the cosmological constant. An encouraging hint is that for the Calabi-Yau models, which are related to topological string theory, a toy version of quantum gravity, the entropy does grow much slower (\ref{CYscaling})  than the conventional models. It would be interesting to settle this issue in the future.

\vspace{0.2in} {\leftline {\bf Acknowledgments}}
\nopagebreak

I would like to thank the Mainz Institute for Theoretical Physics (MITP) of the Cluster of Excellence PRISMA$^{+}$ (Project ID 39083149) for its hospitality and support, and the participants of the MITP workshop ``Spectral Theory, Algebraic Geometry, and Strings" especially Jie Gu, Lukas Schimmer for stimulating discussions. I also thank the Bethe Center for Theoretical Physics in Bonn  for hospitality during parts of this work and Albrecht Klemm for comments on the draft. Furthermore, I thank Bao-ning Du, Jun-Hao Li, Gao-fu Ren, Pei-xuan Zeng  for earlier relevant discussions. This work was supported in parts by National Natural Science Foundation of China (Grant No. 12247103).

\appendix

\addcontentsline{toc}{section}{References}


\providecommand{\href}[2]{#2}\begingroup\raggedright\endgroup

\end{document}